\documentclass[conference]{IEEEtran}

\usepackage{algorithm}
\usepackage{algorithmic}

\IEEEoverridecommandlockouts
\usepackage{cite}
\usepackage{amsmath,amssymb,amsfonts}
\usepackage{bbm}
\usepackage{graphicx}
\usepackage{textcomp}
\usepackage{xcolor}
\newtheorem{assumption}{Assumption}
\newtheorem{pb}{Problem}

\newtheorem{lemma}{Lemma}

\newtheorem{proposition}{Proposition}

\allowdisplaybreaks
\def\BibTeX{{\rm B\kern-.05em{\sc i\kern-.025em b}\kern-.08em
    T\kern-.1667em\lower.7ex\hbox{E}\kern-.125emX}}
\graphicspath{{figures/}}
\begin{document}

\title{Age of Information Aware Cache Updating\\ with File- and Age-Dependent Update Durations}
\author{
   \IEEEauthorblockN{Haoyue~Tang\textsuperscript{1}, Philippe Ciblat\textsuperscript{3},~Jintao~Wang\textsuperscript{1,2},~ Mich\`ele Wigger\textsuperscript{3},~Roy Yates\textsuperscript{4}}
\IEEEauthorblockA{
	\textsuperscript{1}Beijing National Research Center for Information Science and Technology (BNRist),\\
	Dept. of Electronic Engineering, Tsinghua University, Beijing, China\\
	\textsuperscript{2}Research Institute of Tsinghua University in Shenzhen, Shenzhen, China\\
	\textsuperscript{3}Telecom ParisTech, Institut Polytechnique de Paris, Paris, France\\
	\textsuperscript{4}Rutgers University, New Brunswick, NJ, USA\\
        \{thy17@mails, wangjintao@\}tsinghua.edu.cn, \{ciblat, wigger\}@telecom-paristech.fr, ryates@rutgers.edu}}
\maketitle
\begin{abstract}
  We consider a system consisting of a library of time-varying files, a server that at all times observes the current version of all files, and a cache that at the beginning stores the current versions of all files but afterwards has to update 
  these files from the server.   Unlike previous  works,  the update duration is not constant  but depends on the file and its  \emph{Age of Information (AoI)}, i.e., the time elapsed since the current version in the cache was created.
 The goal of this work is to design an update policy that minimizes  the average AoI of all  files with respect to a given popularity distribution. 
Actually a relaxed problem, close to the original optimization problem, is solved and a practical update policy is derived. The update policy relies on the file popularity and on the functions that characterize the update durations of the files depending on their AoI. Numerical simulations show a significant improvement of this new update  policy compared to the so-called square-root policy that is optimal under file-independent and constant update durations. 
\end{abstract}

\section{Introduction}
\let\thefootnote\relax\footnotetext{\noindent -----------------\\ This work was supported by the ERC (Grant No. 715111), the National Key R\&D Program of China (Grant No.2017YFE0112300), Shenzhen basic Research Project (No.JCYJ20170816152246879) and the Tsinghua University Tutor Research Fund.}

Caching, i.e., prestoring popular contents in cache memories close to end users, has become a popular tool to reduce congestion and latency in communication networks. For files that are both time-varying and time-sensitive, i.e., users wish to access recent versions of the files, the files have to be updated regularly. 
The size of an update thereby depends on the original size of the file and on the time elapsed since the latest version in the cache was created, i.e., on the file's \emph{Age of Information (AoI)}. In fact, if a file has been updated recently, then the data has not changed significantly, and its update is   small.  Instead, if a file has not been updated for a long time, then most of it needs to be replaced and the update is large. Our work  accounts for this by letting  the update durations depend on the file and the   file's current AoI. 

Various AoI-related optimization problems have been studied recently,  see e.g. \cite{yates12infocom,yates19tit,kam_17_isit,zhou18GC, talak_2018,igor_ton_2018,boyu_preempt_pub,yates17isit,bastopcu_2019,tang_2019}. 
In this paper, we consider a single-server single-cache system where the server stores the current version of all files and the cache user updates the files in its memory by downloading fresh versions from this server. The goal is to minimize the average AoI of all the files in the cache when the average is taken with respect to a fixed popularity distribution. 
The current work assumes a non-stochastic setup similar to  \cite{yates17isit,bastopcu_2019,tang_2019}  where the server can always access the current  version of all the files. 

The contributions of the paper are as follows. We first formulate the optimization problem of minimizing the average AoI under  AoI-dependent update durations. Then, we slightly relax and simplify  the problem and  solve this simplified version. Inspired by the solution, we propose  a new practical cache update policy that respects all the original constraints. Through numerical simulations, we finally characterize the gain  of this policy over the square-root policy obtained in \cite{yates17isit} for constant update durations. 

The remainder of the paper is organized as follows.  Section \ref{sec:pb} states both the original and the simplified optimization problem. 
Section \ref{sec:sol} solves the relaxed optimization problem  through monotonic optimization and convex optimization theory. Section  \ref{sec:algo} presents a practical downloading policy inspired by the solution of the simplified optimization problem. Section \ref{sec:simus} presents numerical simulations and  Section \ref{sec:ccl} finally concludes the paper.

\section{Problem formulation}\label{sec:pb}

\subsection{System setup}
\begin{figure}[htb]
	\centering
	\includegraphics[width=.49\textwidth]{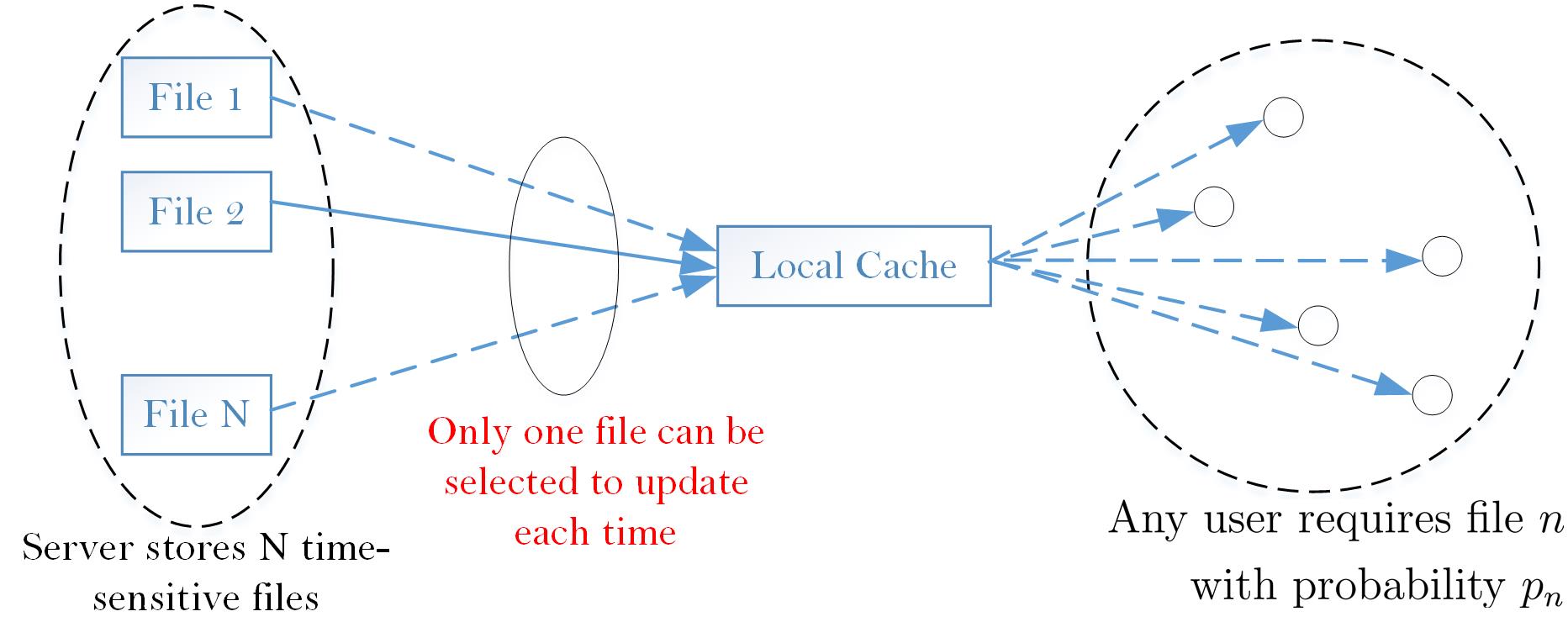}
	\caption{The server-cache-users model.}
	\label{systemmodel}
\end{figure}

The system consists of a remote server and a local server as depicted in Fig. \ref{systemmodel}. 
The remote server has a real-time access to $N$ time-varying and time-sensitive files. The local server starts at time $t=0$ with a fresh version of all files in its cache memory, and given the time-sensitivity of the files it wishes to keep the files as up-to-date as possible.  It will therefore download and update fresh versions of the files from the remote server, where at any given   time  it can only update a single file. It is  also assumed that a new file update can be started only once the previous updates are completed.

 We  consider the observation window  $(0,T]$, for a fixed and given  $T>0$. Let $K_n$, for $n\in\{1,\ldots,N\}$, denote the number of updates of file $n$  in this interval, 
 and  $0< t_{n, 1}<\cdots < t_{n, K_n}$ the \emph{starting times} of these updates. 
 Define then the inter-update intervals
 \begin{equation}\label{eq:tau}
 \tau_{n, i}:=
 \begin{cases}
 t_{n,1}, & i=1;\\
 t_{n, i}-t_{n, i-1},& i=2, \cdots,K_n;\\
 T-t_{n, K_n},& i=K_n+1.
 \end{cases}
 \end{equation}
 Given are also $N$ functions 
  $f_{1}(\cdot),\ldots, f_N(\cdot)$ which describe the time that it takes to update the different files, as will be made more clear in the following. Later in this paper we restrict to specific choices of these functions, but for now we only impose the following assumption:
 	\begin{assumption}[Update function]\label{assumption}
 		Each function $f_n(\cdot)$ is assumed strictly positive, bounded, non-decreasing, concave, and differentiable over $\mathbb{R}_+$. 
 	\end{assumption}

 We now explain the update procedures and the associated age of information (AoI)  process  $\{X_n(t)\}_{t\geq 0}$ of the files.
At time $t=0$ the cache contains fresh versions of all the files, and hence the AoI of each file is  $X_n(0)=0$. The AoI of a given file $n$ then grows as $X_n(t)=t$,
until the cache has \emph{finished} downloading a fresh version of this file. The cache starts the first update of file $n$ at time $t_{n,1}$ and the update is finished at time $t_{n,1} +d_{n,1}$, where $d_{n,1}$ denotes the first update duration of file $n$. This update duration depends on the file's AoI at time $t_{n,1}$ and the update duration function $f_{n}$:
\begin{equation}
d_{n,1}:= f_n(X_{n}(t_{n,1})) = f_n(t_{n,1})=f_n(\tau_{n,1}),
\end{equation}
where the equalities hold because  the AoI at time $t_{n,1}$ is $X_n(t_{n,1})=t_{n,1}$ and by Eq.~\eqref{eq:tau}. 

\begin{figure}[t]
 	\centering
 	\includegraphics[width=.5\textwidth]{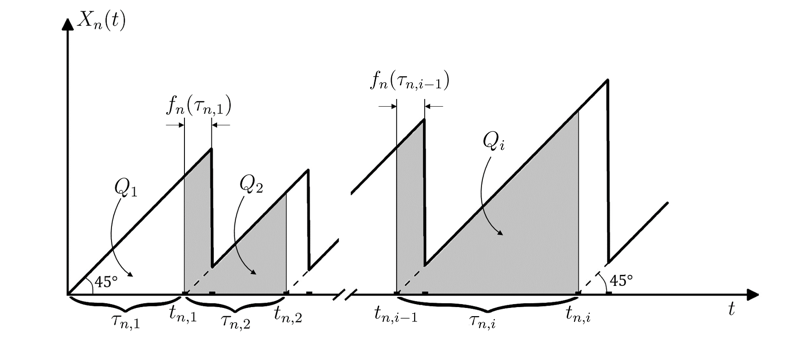}
 	\caption{A sample path of AoI evolution $X_n(t)$ for file $n$.}
 	\label{AoIevlove}
 \end{figure}

At the \emph{end} of this first update, at time $t=t_{n,1}+d_{n,1}$, the AoI of file $n$ drops to the time elapsed since the fresh version was created, i.e., to $d_{n,1}$, and then 
grows again as $X_n(t)=t-t_{n,1}$ until the cache has \emph{finished} the second update of the file.  This second update starts at time $t_{n,2}$ and finishes at time $t_{n,2}+d_{n,2} = t_{n,2}+f_n(\tau_{n,2})$.  
When the second update ends, i.e., at time $t=t_{n,2}+d_{n,2}$, the AoI drops to $d_{n,2}$ and then grows as  $X_n(t)=t-t_{n,2}$, and so forth.  A sample path of the AoI is depicted in Fig~\ref{AoIevlove}.  
To summarize, the  AoI of file $n  \in\{1,\ldots,N\}$ is:
\begin{IEEEeqnarray}{rCl}
X_{n}(t) &=& \begin{cases}  t \ , &   t\in\big[0, t_{n,1}+d_{n,1}\big ) ;\\ 
 (t- t_{n,k}) \ , & t\in\big[ t_{n,k}+d_{n,k}, \\
 & \hspace{1.5cm} t_{n,k+1}+d_{n,k+1}\big),
 \end{cases} 
\end{IEEEeqnarray} 
where 
\begin{equation}\label{eq:dn}
d_{n,k}:=f_n(\tau_{n,k}), 
\end{equation}
and the average AoI of a given file $n$ is:
\begin{equation}
\overline{X}_n=\frac{1}{T}\int_{0}^TX_n(t)dt.
\end{equation}
The main focus of this paper is on the expected AoI $\mathbb{X}$ over a file that is randomly chosen from the set $\{1,\ldots, N\}$ according to a given   file popularity distribution $p_1,\ldots, p_N$,
	\begin{equation}\label{eq:expAoI}
\mathbb{X}:=\sum_{n=1}^Np_n\overline{X}_n.
	\end{equation}
The 
update  times $\{t_{n,1},\ldots, t_{n,N}\}_{n=1}^N$ have to ensure that at each point in time only a single file is being updated. This is equivalent to requiring that the intervals $[t_{n,i},t_{n,i}+d_{n,i})$ are disjoint for all  $n\in\{1,\ldots, N\}$ and $i\in\{1,\ldots, K_n\}$.

\subsection{Optimization problem}

The goal is to minimize the expected AoI $\mathbb{X}$ in \eqref{eq:expAoI} over all feasible update times $\{t_{n,1},\ldots, t_{n,K_{n}}\}_{n=1}^N$. We exploit the one-to-one correspondence  in \eqref{eq:tau} and the functional relationship in \eqref{eq:dn} to express the optimization problem in terms of  
 the inter-update intervals $\{\tau_{n,1},\ldots, \tau_{n,K_n+1}\}_n$. Moreover, we simplify the cost function by splitting  the integral in the computation of $\overline{X}_n$  into $K_n+1$ subintervals:
 \begin{equation}
 \overline{X}_n=  \frac{1}{T}\sum_{i=1}^{K_n+1}\int_{t=t_{n,i-1}}^{t_{n,i}} X_n(t) dt =\frac{1}{T}\sum_{i=1}^{K_n+1} Q_{n,i},
 \label{XsumQ}
 \end{equation}
 where 
 \begin{equation}\label{Qarea}
 Q_{n,i}: =\int_{t_{n,i-1}}^{t=t_{n,i}}X_n(t)dt = \tau_{n, i-1}\cdot f_n(\tau_{n, i-1}) +\frac{1}{2}\tau_{n, i}^2.
 \end{equation}
 Here the second equality holds because the integral corresponds to the sum  of a parallelogram and a triangle as depicted by the gray area in Fig.~\ref{AoIevlove}. %

 We reach the following optimization problem.

\begin{pb}[Original problem]
\begin{subequations}
 	\begin{equation}\label{eq:Xbar}
 	\min_{\{K_n,\tau_{n,i}\}} \sum_{n=1}^N p_n\left(\sum_{i=1}^{K_n+1}\frac{1}{2}\tau_{n, i}^2+\sum_{i=1}^{K_n}\tau_{n, i}\cdot f_n(\tau_{n, i})\right)
 	\end{equation} 
where the minimum is over positive integers $K_n$ and positive real numbers $\tau_{n,i}$ that satisfy the following two conditions:
\begin{equation}\label{eq:C1}
\sum_{i=1}^{K_n+1} \tau_{n,i}=T, \quad \forall n\in\{1,\ldots,N\},
\end{equation}
and for all $i,j,n,n'$ satisfying $(n,i)\neq (n',j)$:
\begin{equation}\label{C2}
\left( \sum_{k=1}^i \tau_{n,k} -   \sum_{k=1}^j \tau_{n',k} \right) \notin \Big( - f_{n}(\tau_{n,i}) , \ f_{n'}(\tau_{n',j})\Big].
\end{equation}
        \label{StrictPrimalProblem}
      \end{subequations}
          \end{pb}
 Condition \eqref{C2} holds because  the update intervals $\big[\sum_{k=1}^i \tau_{n,k}, \sum_{k=1}^i \tau_{n,k}+f_{n}(\tau_{n,i})\big)$ and  $\big[\sum_{k=1}^j \tau_{n',k}, \sum_{k=1}^j \tau_{n',k}+f_{n'}(\tau_{n',j})\big)$ have to be disjoint.

     

\section{A Relaxed Suboptimal Solution}\label{sec:sol}
The presented optimization problem seems hard, even in the asymptotic regime $T\to \infty$, on which we will focus shortly. We therefore
relax  constraint   \eqref{C2} into the next constraint
\begin{equation}\label{C2r}
  \sum_{n=1}^N\sum_{i=1}^{K_n}f_n(\tau_{n,i})\leq T,
\end{equation}
i.e., several files can be updated simultaneously as long as the global update duration remains smaller than the observation window. We also choose (possibly suboptimally) uniform inter-update intervals
\begin{equation}
\tau_{n,i}=	\overline{\tau}_{n}:=\frac{T}{K_n+1}.
\label{taukrelationship}
\end{equation}
The optimization problem then consists in finding the optimal choices of $\{\overline{\tau}_{n}\}$, and is stated in Section \ref{subsec:pb}. 
To motivate the choice in \eqref{taukrelationship}, notice that it is optimal when the update durations are constant and identical across files \cite{yates17isit}. Moreover, in Section~\ref{sec:algo} we  present a practical update policy that updates only a single file at each time, and has a performance close to the solution to  our new optimization problem.

\subsection{A Relaxed Suboptimal Problem}\label{subsec:pb}

In what follows, consider the asymptotic regime $T \to \infty$. By \eqref{eq:Xbar}, the expected AoI $\mathbb{X}$ grows without bound unless  for all files $n$ the number of updates $K_n\to \infty$ as $T\to \infty$. We can therefore assume in the following
\begin{equation}
\lim_{T \to \infty}\frac{K_n}{K_n+1}= 1. \label{eq:limKTn}
\end{equation}

Plugging \eqref{taukrelationship} and \eqref{eq:limKTn} into  \eqref{eq:Xbar} and \eqref{C2r} results in the following new optimization problem, which approximates the original problem in the asymptotic regime where $T\to \infty$. 

\begin{pb}[Relaxed suboptimal problem -- version 1]\label{pb3}
  \begin{subequations}
    \begin{equation}
\min_{\{\overline{\tau}_{n}\}_n} \sum_{n=1}^N p_n\left(\frac{1}{2}\overline{\tau}_n+f_n(\overline{\tau}_n)\right)
\label{averageAoI}
\end{equation}
s.t. $\overline{\tau}_{n}\geq 0, \ \forall n$, and
\begin{equation}\label{C2p}
        \sum_{n=1}^N \frac{f_n(\overline{\tau}_{n})}{\overline{\tau}_n}\leq 1.
      \end{equation}
      \end{subequations}
\end{pb}
We reformulate the optimization in terms of the \emph{file utilization ratios}, i.e., the fraction of time that each file is being updated,
\begin{equation}
	\lambda_n:=\frac{f_n(\overline{\tau}_n)}{\overline{\tau}_n}.
	\label{lambdarelationship}
\end{equation}
For finite $T$, this fraction is $(1/T)\sum_{i\in\{1, \cdots, K_n\}}f_n(\tau_{n, i})=K_nf_n(\overline{\tau}_{n})/((K_n+1)\overline{\tau}_n)$, which by \eqref{eq:limKTn} tends to $\frac{f_n(\overline{\tau}_{n})}{\overline{\tau}_n}$ as $T \to \infty$. Due to  Eq. \eqref{C2p}, we get  $\lambda_n\leq 1$ for the feasible points of Problem  \ref{pb3}, i.e., the duration spent for updating the $n$-th file $f_n(\overline{\tau}_n)$ is smaller than the inter-update duration $\overline{\tau}_n$. 

We have the following useful lemma on the function
\begin{equation}\label{eq:g}
  g_n(t):=\frac{f_n(t)}{t}.
  \end{equation}

  \begin{lemma}\label{lem:1}  Under Assumption \ref{assumption}, the function $t\mapsto g_n(t)$ for $t\in\mathbb{R}_+$ is strictly decreasing and its image is $(0, \infty)$. Consequently, $g_n$ has an inverse function denoted by $g_n^{(-1)}$, which is also strictly decreasing.
  \end{lemma}
  \begin{IEEEproof}
  	The  derivative can be upper bounded as:
  	\begin{align}
  	g_n'(t)=\frac{tf_n'(t)-f_n(t)}{t^2}\overset{(a)}{\leq}\frac{-f_n(0)}{t^2}\overset{(b)}{<}0,
  	\end{align}
  	where the inequality $(a)$ is due to the concavity of the function $f_n$ and the inequality $(b)$ is due to its positivity. So $g_n$ is strictly decreasing. Since $\lim_{t\rightarrow 0}f_n(t)=f_n(0)> 0$, we have $\lim_{t\rightarrow 0}\frac{f_n(t)}{t}=\infty$. As the function $f_n$ is upper-bounded, we also have $\lim_{t\rightarrow\infty}\frac{f_n(t)}{t}=0$. Since $f_n$ is differentiable (and so continuous) and strictly decreasing, its image is $(0, \infty)$. The rest of the proof is straightforward.
  \end{IEEEproof}
Therefore, $\bar{\tau}_n= g_n^{(-1)}(\lambda_n)$, and 
 Problem \ref{pb3} is easily  rewritten as an  optimization problem over  $\{\lambda_n\}_{n=1}^N$.
\begin{pb}[Relaxed suboptimal problem -- version 2]\label{pb4}
	Let 
	\begin{equation}
	h_n(\lambda):=g_n^{(-1)}(\lambda)\left(\frac{1}{2}+\lambda\right).
	\end{equation}
Problem 2 is equivalent to:
	\begin{subequations}
		\begin{equation}
		\min_{\{\lambda_n\}_n} \sum_{n=1}^N p_n\cdot h_n(\lambda_n) 
		\label{relaxobj}
		\end{equation}
		s.t. $\lambda_{n}\geq 0, \ \forall n$, and
		\begin{equation}\label{relaxconstraint}
		\sum_{n=1}^N \lambda_n \leq 1.
		\end{equation}
	\end{subequations}
\end{pb}

We finish this subsection by showing that the optimal solution must lie on the boundary of the feasible set. In subsequent subsections we  discuss  numerical optimization methods that can be used to solve our problem. We also present the KKT conditions,  which can be used to simplify the search of the optimal solution when the  function $h_n$ is convex. 

We have the following auxiliary lemma, which will be useful throughout the paper. 

\begin{lemma}\label{prop:decrease}
	The function $h_n(\cdot)$  is strictly decreasing over $\mathbb{R}_+$. \end{lemma}
\begin{IEEEproof}
	According to Proposition \ref{lem:1}, $g_n^{(-1)}(\lambda)$ is strictly decreasing. In addition, thanks to Eq.~\eqref{eq:g}, we get  $f_n(g_n^{(-1)}(\lambda))= g_n^{(-1)}(\lambda)\cdot g_n(g_n^{(-1)}(\lambda))=g_n^{(-1)}(\lambda)\lambda$. Since $f_n$ is non decreasing and $g_n^{(-1)}$ is strictly decreasing, the composition $f_n \circ g_n^{(-1)}$ is a non increasing function. 
\end{IEEEproof}

\begin{proposition}\label{eq:equal}
  Let $\{\lambda_n^\star\}_n$ be the optimal solution of Problem \ref{pb4}. It satisfies:
  \begin{equation}\label{eq:C3}
    \sum_{n=1}^N\lambda_n^\star=1,
  \end{equation}
i.e., it  lies on the boundary of the feasible set.
  
\end{proposition}
\begin{IEEEproof}
  By contradiction, let us assume $\sum_{n=1}^N\lambda_n^\star< 1$. Then for an arbitrary $n_0$, we replace $\lambda_{n_0}^\star$ with $\lambda_{n_0}^\star+ \delta_{n_0}$ to force equality in Eq.~\eqref{eq:C3}. As $h_n$ is strictly decreasing, we get $h_n(\lambda_{n_0}^\star+ \delta_{n_0})< h_n(\lambda_{n_0}^\star)$. And the point $\lambda_{n_0}^\dag=\lambda_{n_0}^\star+ \delta_{n_0}$  and  $\lambda_{n}^\dag=\lambda_{n}^\star$ for $n\neq n_0$ is better than the optimal one, which concludes the proof.
\end{IEEEproof}

\subsection{Monotonic optimization solution}\label{sec:opt}
Problem \ref{pb4} can be cast into the monotonic optimization framework \cite{jor_2010,bjorn_2012,bjorn_2013} because the constraints are linear (and thus convex) and  the cost function 
is strictly decreasing by Lemma \ref{prop:decrease}. The optimal solution can thus be found using  the so-called Branch-Reduce-Bound (BRB)  \cite{bjorn_2013}. Notice that the function $h_n(\lambda)$ grows without bound when $\lambda\to 0$ (In this limit the file is not updated and its age diverges, see Eq.~\eqref{eq:g1_lambert}.). The  BRB algorithm therefore  has to remove a tiny neighborhood around the origin from the initially selected box.

\subsection{KKT based algorithm}
The function $h_n$ is determined by the update function $f_n$ and is generally not convex. This makes that in general the Karush-Kuhn-Tucker (KKT) conditions are only necessary but not sufficient for an optimal solution. However, for many  practically relevant choices of the update function $f_n$ (see Section \ref{subsec:fn} for more details), the function $h_n$ is convex. We therefore derive the KKT conditions in this subsection.

Let us define the Lagrangian  
$$\mathcal{L}(\lambda_1,\dots, \lambda_N,\nu)=\sum_{n=1}^N h_n(\lambda_n) +\nu \left(\sum_{n=1}^N \lambda_k - 1\right) $$
with $\nu\geq 0$ the Lagrange multiplier. The KKT conditions then state that the primal-dual optimal solutions $(\lambda_1^\star, \cdots, \lambda_N^\star, \nu^\star)$ must satisfy
\begin{eqnarray} 
 p_n h'_n(\lambda_n^\star)+\nu^\star &=&0, \quad \forall \ n\label{eq:KKT1}\\
\nu^\star\left(\sum_{n=1}^N \lambda_n^\star - 1\right) &=&0,\label{eq:slack}
\end{eqnarray}
where $h_n'$ denotes the first-order derivative of $h_n$:
\begin{equation}\label{eq:kktB}
h_n'(\lambda):={g_n^{(-1)}}'(\lambda)\left(\frac{1}{2}+\lambda\right)+g_n^{-1}(\lambda).
\end{equation} Notice that $h_n'$ is invertible, and the image of this inverse $h_n'^{(-1)}$ is the set of all nonpositive   real numbers $(-\infty,0]$. This latter property holds because $h_n$ is differentially continuous and goes from $+\infty$ to $0$.
We can thus rewrite \eqref{eq:KKT1} as
\begin{equation}\label{eq:kkt}
  \lambda_n^\star= h_n'^{(-1)}\left(-\frac{\nu^\star}{p_n}\right), \quad \forall n.
\end{equation}
Condition \eqref{eq:slack} is subsumed by the stronger condition  $\sum_{n=1}^N\lambda_n^\star=1$, which was proved in 
Proposition~\ref{eq:equal}. The  optimal primal variables $\lambda_1^\star ,\ldots, \lambda_N^\star$ are thus given by \eqref{eq:kkt} for some ``waterlevel" $\nu^\star\geq 0$  that needs to be chosen so that $\sum_{n=1}^N\lambda_n^\star=1$.
Certain  functions $f_n$ permit to find a closed-form expression for $h_n'^{(-1)}$. For other functions one needs to 
search over the entries of  a Look Up Table  to find the desired values of 
 $h_n'^{(-1)}$. 

Sometimes it is more convenient to perform a change of variables and express the KKT conditions in terms of the optimal inter-update intervals $\overline{\tau}_n^\star={g_n^{(-1)}}(\lambda_n^\star)$.  Since, ${g_n^{(-1)}}'(\lambda_n^\star)=\frac{1}{g_n'(\tau_n^\star)}=\frac{(\overline{\tau}_n^\star)^2}{f_n'(\overline{\tau}_n^\star)\overline{\tau}_n^\star-f_n(\overline{\tau}_n^\star)}$, Eq.~\eqref{eq:kkt} is 
	equivalent to 
\begin{equation}\label{eq:kktC}
  \overline{\tau}_n^\star + \frac{( \overline{\tau}_n^\star)^2}{ f_n'( \overline{\tau}_n^\star) \overline{\tau}_n^\star -f_n( \overline{\tau}_n^\star)} \left(\frac{1}{2}+ \frac{f_n( \overline{\tau}_n^\star)}{ \overline{\tau}_n^\star}  \right) =  -\frac{\nu^\star}{p_n}, \forall n,
\end{equation}
where 
 $f_n'$ denotes the derivative  of $f_n$.  This equation can be easier to solve because it does not include 
 the inverse $g_n^{(-1)}$. For instance, for $f_n(t)= B_n-(B_n-\varepsilon_n)/(1+t)$ (with $B_n>\varepsilon_n>0$ well tuned in order to ensure the convexity of $h_n$), solving  Eq~\eqref{eq:kktC} is equivalent to finding the positive real-valued root of a fourth-order polynomial and thus a closed-form solution exists.

\section{A practical example for $f_n$}\label{subsec:fn}


In this section, we motivate a specific choice of $f_n$, which  will extensively be used in the simulation part. The idea is that in each unit of time, a certain portion of each file becomes obsolete, and that  the cache and the server know the obsolete parts. These bits can thus be modeled as erasures, and we model the evolution of file $n$ as passing each bit through a Binary Erasure Channel (BEC) with parameter $\Delta_n$. Assume that the  file $n$ initially consists of  $B_n$ bits. Then,  after $t$ time units without update,  any given bit of the file  $n$ undergoes  $t$ sequential applications of a BEC with parameter $\Delta_n$. This transition can be modeled as a BEC with parameter $1-(1-\Delta_n)^t$, and the average number of erased positions in the file after $t$ time units is  $B_n(1-(1-\Delta_n)^t)$.

However, when  $\lim_{t\to 0} f_n(t)=0$, then  degenerate solutions like $\bar{\tau}_n^\star =0$ could be optimal, which is not feasible in practice. We therefore add an offset $\varepsilon_n$ to each update function. Combined with the arguments in the previous paragraph, we  obtain $f_n(t)= B_n - (B_n-\varepsilon_n)(1-\Delta_n)^t$  or expressed in an exponential form:
\begin{equation}\label{eq:fn}
f_n(t)= B_n - (B_n-\varepsilon_n)e^{-\beta_n t}
\end{equation}
with $\beta_n= - \log(1-\Delta_n)$. Notice that $\beta_n>0$ and that such a function $f_n$ always satisfies Assumption \ref{assumption}.

For the choice in Eq.~\eqref{eq:fn},
\begin{equation}\label{eq:g1_lambert}
g_n^{(-1)}(\lambda)=\frac{B_n}{\lambda}+\frac{1}{\beta_n}W\left(-\frac{\beta_n(B_n-\varepsilon_n)}{\lambda}e^{-\frac{\beta_nB_n}{\lambda}}\right),
\end{equation}
where $W(\cdot)$ denotes the Lambert function.
Notice that the associated function  $h_n=g_n^{(-1)}(\lambda)(\frac{1}{2}+\lambda)$ is  convex for certain values of $\epsilon_n, B_n, \beta_n$ (for instance, for  the set of parameters selected in Section \ref{sec:simus}). In this case the solution can be found based on the KKT conditions.  For other values of $\epsilon_n,B_n, \beta_n$ (for instance, for $B_n=1$, $\varepsilon_n=0.02$, and $\beta_n=10$) the function $h_n$ is non-convex and we suggest to use the BRB algorithm to find the optimal solution for the relaxed problem. %

\section{A Practical Scheduling Algorithm}\label{sec:algo}
The question now is: how to apply the result of the previous sections to obtain a practical scheduling algorithm that satisfies all the original constraints? In particular, only a single update should be scheduled at any given point in time, and a new update can only start once  the previous update has terminated.

To describe the practical update algorithm, let $\{{\lambda}^\star_n\}$ be an optimal solution to Problem \ref{pb4}, which is either obtained with the BRB algorithm or with the KKT-based algorithm (if $h_n$ is convex). Then set  $\overline{\tau}_n^\star:=g_n^{(-1)}(\lambda_n^\star)$.
If the algorithm has to schedule a new update at a given time $t$ (because the previous file update just finished), it will choose the file that is currently most urgent, i.e., whose AoI is closest to its maximum target AoI $\bar{\tau}_n^\star$. More precisely,  the algorithm  schedules any of the files  $n_0(t)\in \{1,\ldots, N\}$ that satisfies
$$n_0(t)=\arg\min_{n\in\{1,\cdots, N\}} \overline{\tau}_n^\star-X_n(t). $$ 

\section{Simulations}\label{sec:simus}
In this Section, we numerically compare our idealized and practical scheduling policies with  the  so-called square-root (sqrt) strategy developed in \cite{yates17isit}, which is optimal when the update duration equals the same constant value for all files.\footnote{The work in \cite{yates17isit} considered update ratios  as optimization parameters and not utilization ratios. For constant update durations the two notions coincide.}  Under constant update durations,  the optimal utilization ratios $\{\lambda^\star_n\}$ can be determined analytically (see also \eqref{eq:sqrt_w} below for the special case $B_1=\ldots=B_N$):
\begin{equation}\label{eq:srqt}
\lambda_n^\star=  \lambda_n^{\text{sqrt}}=\frac{\sqrt{p_n}}{\sum_{i=1}^N\sqrt{p_i}}.
  \end{equation}


We present numerical simulations for two choices of the update functions: {\it i)} $f_n(t)=B_n$, and {\it ii)} $f_n(t )$ given in Eq. \eqref{eq:fn}. In all the following figures, blue curves indicate the performance under constant identical update durations and orange curves the performance under one of the two choices of functions $\{f_n$\}. Solid curves indicate the solutions of the relaxed problems 
(either Problem \ref{pb4} for the proposed scheduling policy or the optimization problem described in \cite{yates17isit}) and dashed curves correspond to the proposed practical scheduling algorithms (see Section \ref{sec:algo}) that  avoid collisions. 


\subsection{File-dependent but age-independent update durations}
Assume $f_n(t)=B_n$. In this case, $g_n(t)=B_n/t$ and  $g_n^{(-1)}(\lambda)=B_n/\lambda$ leading to:
$$h_n(\lambda)= \frac{B_n}{2\lambda}+B_n.$$
This function is convex, and we just need to solve the KKT conditions. According to Eq.~\eqref{eq:kkt}, we obtain 
\begin{equation}\label{eq:sqrt_w}
  \lambda_n^\star= \frac{\sqrt{p_nB_n}}{\sum_{i=1}^N \sqrt{p_iB_i}}.
\end{equation}
Notice that the policy given in Eq.~\eqref{eq:sqrt_w} is a slight modification of the one given in Eq.~\eqref{eq:srqt} by weighting the popularity of a file  with its update duration.
According to Eq.~\eqref{eq:sqrt_w}, for two files having the same update duration, the most popular one will be updated more often. For two files with the same popularity, the file with longer update duration will get a larger utilization ratio. However, as the update ratio (proportion of updates done within the observation window) is equal to $\lambda_n/B_n$, the files with longer update duration will be updated less frequently. 

We split the files into two categories: for $n\in\{1, 2, \cdots, N/2\}$, $B_n=1$ with popularity  $p_n=\propto 1/n^\alpha$; for  $n\in\{N/2+1, 2, \cdots, N\}$, $B_n=5$ and $p_n=p_{n-N/2} \propto 1/(n-N/2)^\alpha$. Each category thus obeys a Zipf-like distribution with parameter $\alpha$. We fix $\alpha=1.8$. 

\begin{figure}[h]
	\centering
	\includegraphics[width=.46\textwidth]{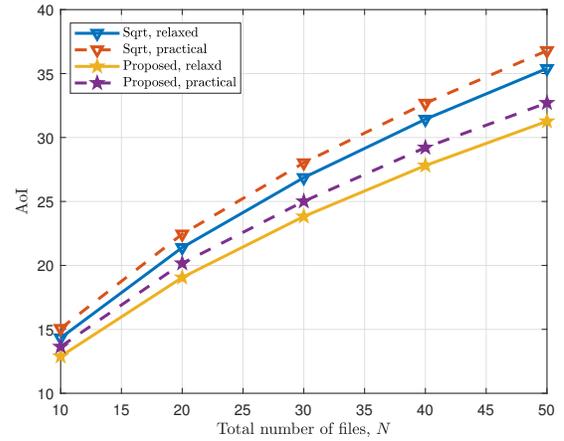}
	\caption{Average AoI versus $N$ (non-identical but constant update durations).}
	\label{fixed}
\end{figure}

In Fig. \ref{fixed}, we plot the average AoI versus the number of files $N$. We observe that the proposed strategy outperforms the square root law based strategy. For instance, the gain is $10\%$ at $N=50$.
Moreover the loss in performance of the practical algorithm (which prevents from collision, i.e., only one file is scheduled) is small compared to the  relaxed solution (which does not prevent to schedule multiple files).


\subsection{Age-dependent update durations}
Throughout this section, assume $f_n$ as in Eq.~\eqref{eq:fn} and fix  $\varepsilon_n=0.02, B_n=1$, for all files $n$. We will consider different values for the parameters $\beta_n$. In particular, we consider setups where all $\beta_n$s are the same, and thus the update durations depend only on a file's age but not on the identity (index) of the file, and setups with different $\beta_n$s.  Throughout this section, we assume that the popularity of the files follows a Zipf-distribution with  parameter  $\alpha=1.8$, 
and we apply the BRB algorithm as described in Section \ref{sec:opt} to find the optimal solution.

\begin{figure}[h]
	\centering
	\includegraphics[width=.44\textwidth]{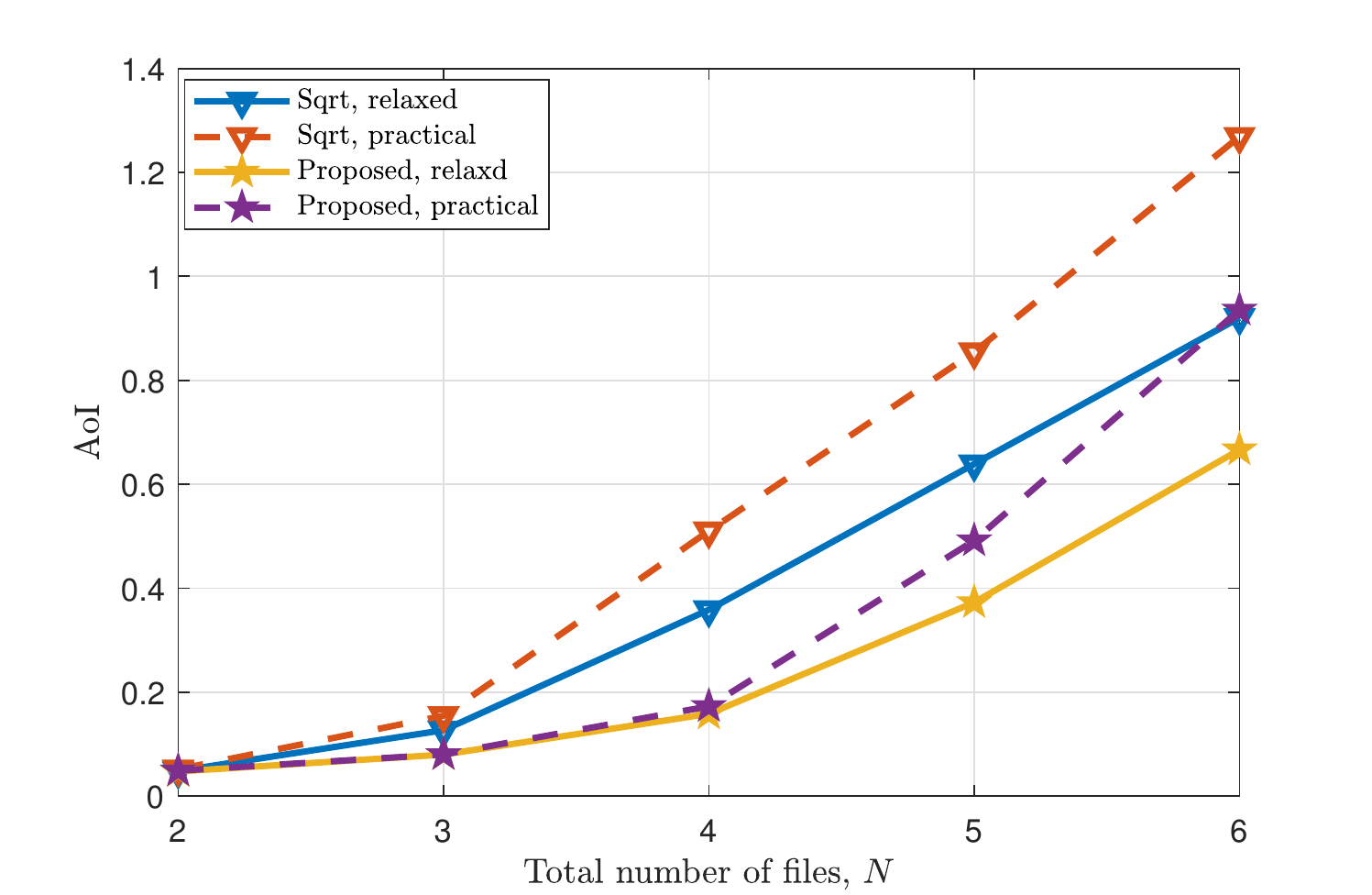}
	\caption{Average AoI versus $N$ (non-constant update durations).}
	\label{exponentialN}
\end{figure}

In Fig. \ref{exponentialN} we consider the same setup with $\beta_n=0.015$, $\forall n$, but we plot the average AoI versus $N$ when the update  function is identical for all files. The proposed strategy achieves a smaller AoI compared to  the square root law. For $N=5$, the gain is around $50\%$ for the practical algorithm. 

\begin{figure}[h]
	\centering
	\includegraphics[width=.44\textwidth]{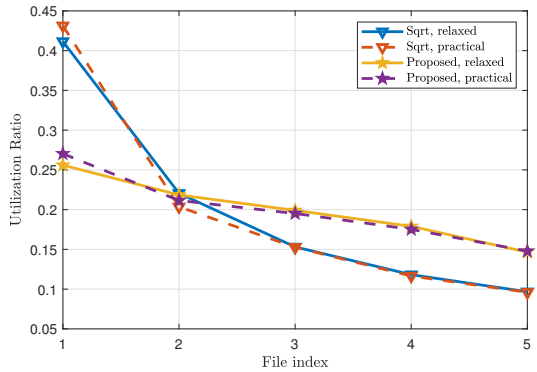}
	\caption{Utilization ratio versus file index (non-constant update size).}
	\label{exponentialsqrt}
\end{figure}

In Fig. \ref{exponentialsqrt}, we plot the individual utilization ratios versus the file indices (sorted by popularity's order) when $N=5$. Once again, $\beta_n=0.015$, $\forall n$. We observe that the optimal utilization ratio  significantly differs from the square root law. Actually, the most popular files are updated more frequently and  their update durations are shorter. This finally leads to a smaller utilization ratio for the most popular files.

\begin{figure}[h]
	\centering
	\includegraphics[width=.46\textwidth]{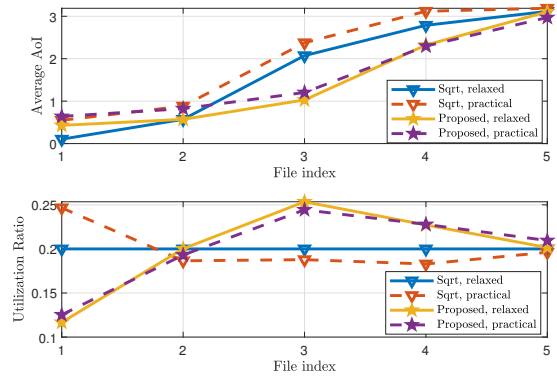}
	\caption{Utilization ratio (top) and average AoI (bottom) versus file index (non-identical and non-constant update durations).}
	\label{identicalUR}
\end{figure}

In Fig. \ref{identicalUR}, we plot the individual utilization ratio and the average AoI versus the file index for $N=5$ when all files have same  popularity ($p_n=1/N$, $\forall n$) but different update duration parameter $\beta_n =0.1\times n$, $\forall n$. 
 Notice that for large values of $n$, the parameter $\beta_n$ is also large and the update function $f_n$ in Eq.~\eqref{eq:fn} is almost constant. For these files, the utilization ratios of the proposed algorithms are  close to the ones under the square root law (which is optimal under constant update durations). Instead for small $n$ the parameter $\beta_n$ is small and the update function $f_n$ is strictly increasing for small AoIs. For these files the utilization ratios of our algorithms are significantly smaller than under the square-root law. A closer inspection of our simulations reveals that  these files are updated frequently, but each update is short. For moderate $n$  the  
 utilization ratio is large because these are updated frequently and each update is not very short.

\section{Conclusion}\label{sec:ccl}
This paper proposes a practical algorithm for scheduling updates from a remote server to a local cache when the update duration depends on the file's AoI. The proposed algorithm is shown to have small performance loss compared to the optimal scheduling policy of a relaxed problem. In all these results, a given file popularity is taken into account. 

\bibliography{bibfile}

\end{document}